\documentstyle[11pt,aaspp4]{article}

\def\be{\begin{equation}}
\def\ee{\end{equation}}
\def\go{\mathrel{\raise.3ex\hbox{$>$}\mkern-14mu
             \lower0.6ex\hbox{$\sim$}}}
\def\lo{\mathrel{\raise.3ex\hbox{$<$}\mkern-14mu
             \lower0.6ex\hbox{$\sim$}}}
\def\abs{{({\rm abs})}}
\def\sc{{({\rm sc})}}
\def\hato{{\hat{\bf\Omega}}}

\def\bF{{\bf F}}
\def\br{{\bf r}}

\def\nue{{\nu_e}}
\def\bnue{{\bar\nu_e}}

\begin{document}

\title{Neutrino Transport in Strongly Magnetized Proto-Neutron
Stars and the Origin of Pulsar Kicks:\\ 
The Effect of Asymmetric Magnetic Field Topology}

\author{Dong Lai}
\affil{Department of Astronomy, Space Sciences Building,
Cornell University, Ithaca, NY 14853\\
E-mail: dong@spacenet.tn.cornell.edu}
\and
\author{Yong-Zhong Qian}
\affil{Physics Department, 161-33, California Institute of
Technology, Pasadena, CA 91125}

\begin{abstract}
In proto-neutron stars with strong magnetic fields, the cross
section for $\nu_e$ ($\bar\nu_e$) absorption on neutrons
(protons) depends on the local magnetic field strength due to
the quantization of energy levels for the $e^-$ ($e^+$) 
produced in the final state. If the neutron star possesses
an asymmetric magnetic field topology in the sense that
the magnitude of magnetic field in the north pole is different
from that in the south pole, then asymmetric neutrino emission
may be generated. We calculate the absorption 
cross sections of $\nue$ and $\bnue$ in strong magnetic fields as a 
function of the neutrino energy.  These cross sections exhibit
oscillatory behaviors which occur because new Landau levels for 
the $e^-$ ($e^+$) become accessible as the neutrino energy increases.
By evaluating the appropriately averaged
neutrino opacities, we demonstrate that the change in the local neutrino flux 
due to the modified opacities is rather small. To generate appreciable kick
velocity ($\sim 300$~km~s$^{-1}$) to the newly-formed neutron star, 
the difference in the field strengths at the two opposite poles of the star
must be at least $10^{16}$~G.
We also consider the magnetic field
effect on the spectral neutrino energy fluxes. The oscillatory features
in the absorption opacities give rise to modulations in the
emergent spectra of $\nu_e$ and $\bar\nu_e$.

\end{abstract}

\keywords{stars: neutron -- pulsars: general -- supernovae: general
-- dense matter -- magnetic fields -- radiation transfer}

\section{Introduction}

It has been recognized shortly after the discovery of pulsars
that neutron stars have velocities much in excess of those
of any other normal stellar populations in our Galaxy
(Gunn \& Ostriker 1970; Minkowski 1970).
However, it is only in the last few years that a significant body of
evidence has come into place to support the view that type II supernovae
are nonspherical and neutron stars receive large kick velocities
at birth. Lyne and Lorimer (1994) analyzed
pulsar velocities in light of new proper-motion measurements
(Harrison et al.~1993) and up-to-date pulsar distance scale
(Taylor \& Cordes 1993), and concluded that pulsars were born with a mean
speed of $\sim 450$~km~s$^{-1}$, much larger than previously thought.
More recent studies of pulsar velocities have adopted more sophisticated
satistical methods and included better treatment of selection effects
and uncertainties (Lorimer et al.~1997; Hansen \& Phinney 1997;
Cordes \& Chernoff 1997). They all yielded a result for the pulsar birth
velocity in qualitative agreement (although not in quantitative details)
with that of Lyne \& Lorimer (1994). In particular, Cordes and Chernoff
(1997) found that the three-dimensional space velocities of their sample
of 47 pulsars have a bimodal distribution, with characteristic speeds of
$180$~km~s$^{-1}$ and $700$~km~s$^{-1}$ (corresponding to 
$80\%$ and $20\%$
of the population). They also estimated that pulsars with velocities
greater than $1000$~km~s$^{-1}$ may be underrepresented 
owing to selection effects in pulsar surveys [The uncertainty in the
high-velocity end of the pulsar velocity distribution function has also be
emphasized by Hansen \& Phinney (1997)]. Concrete evidence for the existence of
pulsars with velocities of $\go 1000$~km~s$^{-1}$
has come from the observation of
the Guitar Nebula pulsar (B2224+65), which produces a bow shock when
plowing through the interstellar medium (Cordes, Romani \& Lundgren 1993).
In addition, studies of pulsar-supernova remnant associations
have uncovered a number of pulsars having velocities
much greater than $1000$~km~s$^{-1}$ (Frail et al.~1994), although
in many cases the associations are not completely secure.

Compelling evidence for supernova asymmetry and pulsar kicks
also comes from the detection of geodetic precession in the binary pulsar
PSR 1913+16 (Cordes et al.~1990; Arzoumanian et al.~1996),
and the orbital plane precession in the PSR J0045-7319/B star binary 
(Lai et al.~1995; Kaspi et al.~1996) 
and its fast orbital decay (which indicates
retrograde rotation of the B star with respect to the orbit; see Lai 1996).
These results demonstrate that binary break-up (as originally suggested
by Gott, Gunn \& Ostriker 1970; see Iben \& Tutukov 1996) can not be
{\it solely} responsible for the observed pulsar velocities, and that
{\it natal kicks are required}. In addition, evolutionary studies
of neutron star binary population imply the existence of pulsar kicks (e.g.,
Deway \& Cordes 1987; Fryer \& Kalogera 1997;
Fryer et al.~1998; see also Brandt \& Podsiadlowski 1995).
Finally, there are a large number of direct observations of nearby
supernovae (e.g., Cropper et al.~1988;
Trammell et al.~1993; McCray 1993; Utrobin et al.~1995) and supernova
remnants (e.g., Morse, Winkler \& Kirshner 1995; Aschenbach et al.~1995)
in radio, optical, and X-ray bands which support the notion that supernova
explosions are not spherically symmetric.

The origin of the pulsar velocities is still unknown.
Two classes of mechanisms for the {\it natal kicks}
have been suggested\footnote{These exclude the slow, post-explosion
rocket effect due to electromagnetic radiation from off-centered
magnetic dipole moment (Harrison \& Tademaru 1975).}.
The first class relies on convective instabilities in the
collapsed stellar core and within the rebounding shock
(e.g., Burrows \& Fryxell 1992; Burrows et al.~1995;
Janka \& M\"uller 1994,~1996; Herant et al.~1994).
The asymmetries in the matter and temperature distributions associated with
the instabilities naturally lead to
asymmetric matter ejection and/or asymmetric neutrino emission.
Numerical simulations indicate that the local, post-collapse
instabilities are not adequate to account for kick velocities higher than
a few hundred km~s$^{-1}$ 
(Janka \& M\"uller 1994).
A variant of this class of models therefore relies on the global asymmetric
perturbations seeded in the presupernova cores (Goldreich et al.~1996; see also
Bazan \& Arnett 1994). Clearly, the magnitude of kick velocity depends on
the degree of initial asymmetry in the imploding core (Burrows \& Hayes 1996).
Due to various uncertainties in the presupernova stellar models, it is not
clear at this point whether sufficiently large initial perturbations can be
produced in the precollapse core (Lai \& Goldreich 1998).

In this paper we focus on the second class of models in which the large
pulsar kick velocities arise from asymmetric neutrino emission induced by
strong magnetic fields.
Since $99\%$ of the neutron star binding energy (a few times $10^{53}$~erg)
is released in neutrinos, tapping the neutrino energy would appear to be
an efficient means to kick the newly-formed neutron star.
Magnetic fields are naturally invoked to break the spherical symmetry in
neutrino emission. But the actual mechanism is unclear.
We first briefly comment on previous works on this
and related subjects.

\subsection{Previous Works on Asymmetric Neutrino Emission Induced by
Magnetic Fields}

A number of authors have noted that parity violation in weak
interactions may lead to asymmetric neutrino emission from
proto-neutron stars (Chugai 1984; Dorofeev et al.~1985;
Vilenkin 1995; Horowitz \& Piekarewicz 1997). However, their studies
are largely unsatisfactory for a number of reasons: they either
failed to identify the most relevant neutrino emission/interaction
processes or the relevant physical conditions in proto-neutron stars, or
stopped at estimating the magnetic field effects on neutrino opacities.
Chugai (1984) and Vilenkin (1995) (see also Bezchastnov \& Haensel 1996)
considered neutrino-electron scattering and concluded that the effect is
extremely small\footnote{Note that Chugai's estimate for the electron
polarization in the relativistic and degenerate regime (the relevant
physical regime) is incorrect. This error leads to an overestimate
of the effect as compared to Vilenkin's result.}
(e.g., to obtain $V_{\rm kick}=300$~km~s$^{-1}$ would require
a magnetic field of at least $10^{16}$~G).
However, neutrino-electron scattering is less important
than neutrino-nucleon scattering in determining the
characteristics of neutrino transport in proto-neutron stars.
Similarly, Dorofeev et al.~(1985) considered neutrino emission by Urca
processes in strong magnetic fields. But as shown by Lai \& Qian (1998),
in the bulk interior of the neutron star, the asymmetry associated with
neutrino emission is cancelled by the asymmetry associated with
neutrino absorption. 

Concerning the parity violation effect in proto-neutron stars,
Horowitz \& Li (1997) recently suggested
that the asymmetry in neutrino emission
may be enhanced due to multiple scatterings of
neutrinos by nucleons which are slightly polarized by the magnetic field.
They estimated that a field strength of a few times $10^{12}$~G
is adequate to account for kick velocities of a few hundred km~s$^{-1}$.
However, the paper by Horowitz \& Li (1997) only discusses
the idealized situations of scattering media, and ignores
various essential physics that is needed in a proper treatment of
neutrino transport in proto-neutron stars. In particular, it does not
consider the effect of neutrino absorption which one might suspect to
wash out the cumulative effect from multiple scatterings.
Our own study of the parity violation effect (Lai \& Qian 1998) was
flawed in the treatment of scattering terms in the neutrino transport 
equation. As shown by Arras \& Lai (1998), detailed balance requires
that there be no cumulative effect from multiple scatterings in the
bulk interior of the proto-neutron star where local thermodynamic
equilibrium applies to a good approximation. Enhancement
of neutrino emission asymmetry from multiple scatterings obtains
only after neutrinos thermally decouple from proto-neutron star matter,
and therefore is insignificant.

Bisnovatyi-Kogan (1993) attributed pulsar kicks to
asymmetric magnetic field distribution in proto-neutron stars.
Using the fact that neutron decay rate can be modified by the
magnetic field, he inferred that neutrino emissions from opposite
sides of the neutron star surface are different.
However, neutron decay is not directly relevant for neutrino emission
from a newly-formed neutron star.
Roulet (1997) (whose paper was posted while this paper was being written)
considered the relevant neutrino absorption processes.
But his study was restricted to calculating the neutrino
cross sections as functions of neutrino energy (corresponding to
\S2.1 of this paper), and therefore was insufficient for addressing the
issue of asymmetric neutrino emission.

Finally, it is worthwhile to mention a speculative idea on pulsar kicks
which relies on nonstandard neutrino physics. Kusenko \& Segr\`e (1996)
suggested that asymmetric $\nu_\tau$ emission could result from
the Mikheyev-Smirnov-Wolfenstein flavor transformation
between $\nu_\tau$ and $\nu_e$ inside a magnetized proto-neutron star
because a magnetic field changes the resonance condition for the
flavor transformation.
With a $\nu_\tau$ mass of
$\sim 100$~eV, they claimed that magnetic fields of
$\sim 3\times 10^{14}$~G can give the pulsar a kick velocity of a few hundred
km~s$^{-1}$. However, their treatment of neutrino transport
was oversimplified: e.g., they ignored that neutrinos of different
energies have different resonance surfaces. Furthermore, a simple
geometric effect and realistic proto-neutron star conditions
easily reduce their estimated kick velocity by an order of magnitude.
Most likely, a magnetic field strength of $\sim 10^{16}$~G is needed
to produce the observed average pulsar kick velocity via their mechanism
(see Qian 1997). Similarly strong magnetic fields are required in
variants of the Kusenko \& Segr\`e mechanism, e.g., those considered
by Akhmedov et al.~(1997) (which relies on both the neutrino mass and the
neutrino magnetic moment to facilitate the flavor transformation). 

\subsection{Plan of This Paper}

As should be clear from the brief review in \S1.1,
in spite of quite a number of suggestions, there is at present no
consensus on the magnitudes of the magnetic field induced asymmetry
in neutrino emission from proto-neutron stars and the resulting kick
velocities.
In this paper we study
the effect of asymmetric magnetic field topology on pulsar kicks.
Since the energy levels of $e^-$ and $e^+$ in a magnetic field
are quantized, the $\nu_e$ and $\bnue$ absorption opacities
near the neutrinosphere and the neutrino-matter decoupling region depend on the
local magnetic field strength. If the
{\it magnitude} of magnetic field in the north pole is different from
that in the south pole (the field does not need to be ordered),
then asymmetric neutrino flux may be generated. Here
we ignore the effects of magnetic fields on the
proto-neutron star structure through the equation of state --- these are
secondary effects as far as neutrino transport is concerned.

The organization of this paper is as follows. In \S2 we calculate
the $\nu_e$ and $\bar\nu_e$ absorption cross sections as functions of
neutrino energy in strong magnetic fields. Section 3 summarizes
the basic features of neutrino transport near the neutron star surface. 
In \S4 we derive and evaluate the ``Rosseland mean'' neutrino 
opacities which are directly related to the local neutrino flux.
We then consider in \S5 the change in neutrino flux due to the modified
opacities, and estimate the kick velocity resulting from an asymmetric
magnetic field topology. In \S6 we briefly consider how the 
emergent neutrino spectra may be modified by the strong magnetic field. 
In \S7 we present our conclusion.

Unless noted otherwise, we shall use units in which
$\hbar$, $c$, and the Boltzmann constant $k$ are unity.

\section{$\nu_e$ and $\bar\nu_e$ Absorption Cross Sections in
Strong Magnetic Fields}

In this section we consider how magnetic fields modify 
the (total) neutrino absorption 
($\nu_e+n\rightarrow p+e^-$ and $\bar\nu_e+p\rightarrow n+e^+$)
cross sections for specific neutrino energies. 
We present our calculation of the $\nu_e$ absorption cross section
in detail.
The $\bar\nu_e$ absorption cross section can be obtained in a similar manner.

To begin with, it is useful to review various energy scales involved
in the problem. Near the neutrinosphere and the neutrino-matter decoupling
region, the matter density is typically $10^{11}$--$10^{12}$
g$\,$cm$^{-3}$, and the temperature is about several
MeV. The electrons are extremely relativistic and highly 
degenerate, with Fermi energy $E_{F,e}\simeq
51.5\,(Y_e\rho_{12})^{1/3}~$MeV, where $Y_e\sim 0.1$ is the
electron fraction (the number of electrons per nucleon) and
$\rho_{12}=\rho/(10^{12}\,{\rm g}\,{\rm cm}^{-3})$. 
The neutrons and protons have Fermi energies 
$E_{F,n}=1.4\,[(1-Y_e)\rho_{12}]^{2/3}~$MeV and 
$E_{F,p}=1.4\,(Y_e\rho_{12})^{2/3}~$MeV, respectively, 
and thus they are essentially nondegenerate. 

\subsection{The Cross Sections} 

In strong magnetic fields the transverse motion of the electron
is quantized into Landau levels. The electron energy dispersion 
relation is
\be
E_e=\left[p_z^2+m_e^2(1+2 nb)\right]^{1/2},~~~~n=0,~1,~2,~\cdots
\ee
where $p_z$ is the longitudinal momentum (along the magnetic field),
and $n$ is the Landau level index. The dimensionless
field strength $b$ is defined as 
\be
b\equiv {B\over B_c},~~~~~{\rm with}~~B_c={m_e^2c^3\over\hbar e}
=4.414\times 10^{13}~{\rm G}.
\ee
[$B_c$ is the critical field strength defined by equating the cyclotron energy 
$\hbar eB/(m_e c)$ to $m_e c^2$.] The quantization effect on the proton
is extremely small due to the large proton mass, and is neglected throughout
this paper.   

Since the cyclotron radius (characterizing the size of the Landau
wavefunction) $[\hbar c/(eB)]^{1/2}=
[\hbar/(m_e c)]b^{-1/2}=386\,b^{-1/2}$ fm
is much greater than the range of the weak interaction [$\hbar/(m_W c)
\simeq 2.5\times 10^{-3}$ fm, where $m_W\simeq 80$ GeV is the mass of
W boson], one can use the $V-A$ theory with contact
interaction. Locally (at the point of interaction) the Landau
wavefunction resembles a plane wave. Thus we expect that the
spin-averaged matrix element is the same as in the case without
magnetic fields. This expectation has been borne out by
detailed calculations (e.g., Fassio-Canuto 1969; Matese \& O'Connell
1969). Therefore we only need to consider the effect of magnetic field
on the electron phase space. 

Since the neutrino energy (a few to tens of MeV) of interest
is much less than the nucleon mass, we first consider the case where
the nucleon mass is taken to be infinity, an approximation adopted
in the standard zero-field calculations (e.g., Tubbs \& Schramm 1975).
The cross section for $\nu_e + n\rightarrow p + e^-$ can be written as
a sum over the (spatial and spin) states of the electron (per unit volume) 
\be
\sigma^\abs_B(E_\nu)=A\sum_e(1-f_e)\,\delta (E_\nu+Q-E_e),
\ee
where we have defined 
\be 
A\equiv \pi G_F^2 \cos^2\theta_C (c_V^2+3c_A^2),
\ee
and the other symbols have their usual meanings: 
$G_F=(293\,{\rm GeV})^{-2}$ is the universal Fermi constant,
$\cos^2\theta_C=0.95$ refers to the Cabbibo angle, 
$c_V=1$ and $c_A=1.26$ are the coupling
constants, $Q=m_n-m_p=1.293$ MeV is the neutron-proton mass difference, 
$E_\nu$ is the $\nu_e$ energy,
and $(1-f_e)$ is the Pauli
blocking factor. The electron occupation number $f_e$ is
\be
f_e={1\over \exp\left[(E_e-\mu_e)/T\right] +1},
\ee
where $E_e$ is the electron energy,
$\mu_e$ is the electron chemical potential, and $T$ is the
matter temperature. 
For zero field, the sum becomes $2\int d^3p/(2\pi)^3$, and we have
\be
\sigma^\abs_{B=0}(E_\nu)={A\over \pi^2}(E_\nu+Q)
\left[(E_\nu+Q)^2-m_e^2\right]^{1/2}(1-f_e),
\ee
with $E_e=E_\nu+Q$. For $B\neq 0$, knowing that the
degeneracy of the Landau level (per unit area) is
$g_n eB/hc=g_n m_e^2b/(2\pi)$, with $g_n=1$ for $n=0$ and $g_n=2$ for
$n>0$, we have
\be
\sigma^\abs_B(E_\nu)=A
\int_{-\infty}^\infty
{dp_z\over 2\pi}\sum_n g_n{m_e^2b\over 2\pi}(1-f_e)\,
\delta(E_\nu+Q-E_e)=A{m_e^2b\over 2\pi^2}
\sum_{n=0}^{n_{\rm max}} g_n {E_e\over p_{z,n}}(1-f_e),
\label{absb}\ee
where $E_e=E_\nu+Q$, and $p_{z,n}=\left[E_e^2-m_e^2(1+2nb)\right]^{1/2}$.
The upper limit $n_{\rm max}$ for the sum
is the maximum value of $n$ for which $p_{z,n}$ is
meaningfully defined, i.e.,
\be
n_{\rm max}={\rm Int}\left[{(E_\nu+Q)^2-m_e^2\over 2m_e^2b}\right],
\label{nmax}\ee
where Int$[x]$ stands for the integral part of $x$.
Henceforth, $n_{\rm max}$ is similarly defined, and we shall not
explicitly write out the lower and upper limits for the sum. 

Figure 1 depicts $\sigma^\abs_B(E_\nu)$ for $b=100$ and
$\sigma^\abs_{B=0}(E_\nu)$
as functions of $E_\nu$ for typical conditions ($Y_e=0.1$,
$\mu_e=20$ MeV, and $T=3$ MeV)
in a proto-neutron star. Note that ideally, one should
compare $\sigma^\abs_B(E_\nu)$ and $\sigma^\abs_{B=0}(E_\nu)$ at the same
matter density.  
For $\mu_e\simeq E_{F,e}\gg T$ (so that $n_{e^-}\gg n_{e^+}$), 
the matter density
is related to $\mu_e$ via
\begin{eqnarray}
\rho &=&{m_n\over Y_e}{1\over \pi^2}\int_0^\infty\! p^2 dp\,
f_e,~~~~~~~~~~~~~~({\rm for}~ B=0),\\
\rho &=&{m_n\over Y_e}{m_e^2 b\over 2\pi^2}
\int_0^\infty\! dp_z\sum_n g_n\,f_e,~~~~~({\rm for}~ B\neq 0).
\end{eqnarray}
In practice, we have found that for a given $\mu_e$, the 
dependence of $\rho$ on the field strength is rather weak
for the range of parameters of interest in this paper. 
Thus in all of our calculations here and below, we compare 
the cross sections at a given $\mu_e$. A prominent feature
of $\sigma^\abs_B(E_\nu)$ shown in Fig.~1 is the presence of
``resonances,'' which occur when $p_{z,n}=0$ (cf.~Eq.~[\ref{absb}]),
i.e., at the neutrino energies for which the final state electron lies
in a ``stationary'' Landau level (with zero longitudinal momentum). 
These resonances reflect the divergence of the 
density of states of electrons at these stationary Landau levels. 

The cross section for $\bar\nu_e + p\rightarrow n + e^+$ 
can be similarly calculated. One only needs to replace 
$Q$ with $-Q$ and $\mu_e$ with $-\mu_e$ in the expressions for
$\nu_e$ absorption to obtain the desired expressions for $\bar\nu_e$
absorption. The numerical results for $\bar\nu_e$ absorption cross section
are shown in Fig.~2 for the same physical parameters as
used in Fig.~1.

\subsection{Thermal Averaging}

The divergent cross sections at the ``resonances'' shown in Figs.~1 and 2
are clearly unphysical. They are smoothed out by the thermal motion of
the neutron. To the leading order, the width of the resonance is
$(kT/m_n)^{1/2}E_\nu$, 
corresponding to the Doppler shift of the neutrino energy in the
neutron's rest frame. The recoil energy, of order
$E_\nu^2/m_n$, is much less effective in smoothing out the resonances,
and will be neglected. 
With this approximation, the smoothed cross section can be written as
\be
\langle\sigma^\abs_B(E_\nu)\rangle \simeq
{1\over(2\pi)^2}\int\!v_n^2 dv_n\,f_n\int\!d\cos\theta_n
\,\sigma_B^\abs(E_\nu'),
\label{aver}\ee
where $E_\nu'=E_\nu(1-v_n\cos\theta_n)$ and
$f_n=(2\pi m_n/T)^{3/2}\exp(-m_nv_n^2/2T)$ is the velocity distribution
function for the neutron. 
If we replace the factor $(1-f_e)$ [which depends on $(E_\nu'+Q)$]
in $\sigma_B^\abs(E_\nu')$ with the average value
$\langle 1-f_e\rangle$, which can be approximately
taken as $(1-f_e)$ evaluated at $E_e=E_\nu+Q$,
then the angular integral in Eq.~(\ref{aver}) can be carried out 
analytically. For $B=0$, we have
\be
\langle\sigma^\abs_{B=0}(E_\nu)\rangle \simeq
{A\over (2\pi)^2}\langle 1-f_e\rangle
\int\!v_n dv_n\,f_n{1\over 3\pi^2E_\nu}
(E_e^2-m_e^2)^{3/2}\Biggl |^{E_e=E_\nu(1+v_n)+Q}
_{E_e=E_\nu(1-v_n)+Q},
\label{abs0aver}\ee
which is reduced to 
\be
\langle\sigma^\abs_{B=0}(E_\nu)\rangle \simeq
{A\over \pi^2}(E_\nu+Q)^2\langle 1-f_e\rangle
\left[1+\left({E_\nu\over E_\nu+Q}\right)^2{T\over m_n}\right]
\label{abs0aver0}\ee
if we neglect the electron mass $m_e$.
For $B\neq 0$, we find\footnote{To obtain this result, one has to interchange
the sum and the angular integral. This interchange is valid because
$1/(x-x_n)^{1/2}$ has an integrable singularity at $x=x_n$.}
\be
\langle\sigma^\abs_B(E_\nu)\rangle \simeq
{A\over (2\pi)^2}\langle 1-f_e\rangle
\int\!v_n dv_n\,f_n{m_e^2b\over 2\pi^2E_\nu}
\sum_n g_n\left [E_e^2-m_e^2(1+2nb)\right]^{1/2}\Biggl |^{E_e=E_\nu(1+v_n)+Q}
_{E_e=E_\nu(1-v_n)+Q}.
\label{absaver}\ee
We could have approximately included the neutron recoil effect by replacing
$Q$ with $Q-(E_\nu+Q)^2/(2m_n)$ in Eqs.~(\ref{abs0aver})--(\ref{absaver}).
But we choose to leave out this effect since as mentioned previously,
it is ineffective in smoothing out the resonances in $\sigma^\abs_B(E_\nu)$.
We have numerically evaluated Eqs.~(\ref{abs0aver}) and (\ref{absaver}), 
and the results are depicted in Fig.~1 together with the cross sections
before averaging. Similar results for the $\bar\nu_e$ absorption 
cross section are shown in Fig.~2.
As expected, we find that for $B=0$, 
the averaged cross section 
$\langle\sigma^\abs_{B=0}(E_\nu)\rangle$ is almost identical to
$\sigma^\abs_{B=0}(E_\nu)$ (see Eq. [\ref{abs0aver0}]). 
However, for $B\neq 0$,
the thermal averaging is essential in obtaining the smoothed physical 
absorption cross sections. 

\subsection{Limiting Cases}

The oscillatory behaviors of $\langle\sigma^\abs_B(E_\nu)\rangle$
in Figs.~1 and 2 can be demonstrated analytically. 
Using the Poisson summation formula (see e.g., Ziman 1979):
\be
\sum_{n=0}^{\infty}f(n+{1\over 2})=
\sum_{s=-\infty}^{\infty}(-1)^s\int_0^\infty\!f(x)\,e^{2\pi i s x}dx,
\ee
we can rewrite the sum in Eq.~(\ref{absaver}) as 
\begin{eqnarray}
&&m_e^2b \sum_n g_n\left [E_e^2-m_e^2(1+2nb)\right]^{1/2}\nonumber\\
&&={2\over 3}(E_e^2-m_e^2)^{3/2}
+(2m_e^2b)^{3/2}{\rm Re}\sum_{s=1}^\infty
{1-i\over 4\pi i s^{3/2}}\exp\left[2\pi i s \left({E_e^2-m_e^2\over 2m_e^2b}
\right)\right].
\label{summ}\end{eqnarray}
The first term on the right-hand side of Eq.~(\ref{summ})
corresponds to the zero-field result, and the second term (involving the sum
of many oscillatory terms) corresponds to the correction due to the
quantization effect. The sum is clearly convergent, and thus for small $B$,
the averaged cross section $\langle\sigma^\abs_B(E_\nu)\rangle$ is reduced 
to the zero-field value $\langle\sigma^\abs_{B=0}(E_\nu)\rangle$.

For sufficiently large $B$ at a given $E_\nu$, or, equivalently,
for sufficiently small $E_\nu$ at a given $B$, only the ground state 
Landau level is filled by the electron, i.e., $n_{\rm max}=0$
(see Eq.~[\ref{nmax}]). In this limit, 
$\langle\sigma^\abs_B(E_\nu)\rangle>
\langle\sigma^\abs_{B=0}(E_\nu)\rangle/4$, and only the first term in 
Eq.~(\ref{absb}) needs to be retained. We then have
\be
\langle\sigma^\abs_B(E_\nu)\rangle\simeq A\,{m_e^2b\over 2\pi^2}\,
{(E_\nu+Q)\over [(E_\nu+Q)^2-m_e^2]^{1/2}}\,(1-f_e)
\label{sigbapp}\ee
for $\nu_e$ absorption.
A similar expression for $\bar\nu_e$ absorption is obtained by
replacing $Q$ with 
$-Q$ and $\mu_e$ with $-\mu_e$ in Eq. (\ref{sigbapp}).

\subsection{Discussion}

The physical conditions which Figs.~1 and 2 are based on 
($T=3$~MeV, $Y_e=0.1$, and $\mu_e=20$~MeV, corresponding to a matter
density of $\rho=7.2\times 10^{11}$ g~cm$^{-3}$)
are typical of the neutrinosphere of the proto-neutron star. 
In Fig.~3 we plot the $\nu_e$ and $\bar\nu_e$ absorption cross sections 
for a number of slightly different physical parameters, specified
by $b,~\mu_e$ and $T$ (all with $Y_e=0.1$). 
The cross section for elastic neutral-current scattering 
$\nu + N\rightarrow\nu + N$, $\sigma^\sc(E_\nu)\simeq G_F^2E_\nu^2/\pi$ 
(e.g., Tubbs \& Schramm 1975), is also shown for comparison. 
Absorption dominates the transport of $\nu_e$ for high energies,
with the ratio of absorption to scattering cross section per nucleon
$(1-Y_e)\sigma^\abs/\sigma^\sc$ approaching 5 for $E_\nu\gg\mu_e$. For smaller
energies ($E_\nu\lo \mu_e$), absorption is severely 
suppressed by the Pauli blocking of the final electron states,  
and scattering becomes more important. In the case of $\bar\nu_e$, 
Pauli blocking is negligible, and scattering almost always dominates the
transport, with the ratio of scattering to absorption cross section per nucleon
$\sigma^\sc/[Y_e\sigma^\abs]\simeq 2\,(0.1/Y_e)$ for $E_\nu\go 10$~MeV.

As we can see from Figs.~1-3, for a given set of physical parameters 
of the medium ($T,~Y_e,~\mu_e$ or $\rho$), the magnetic field effect
on the absorption opacity is more prominent for lower-energy 
neutrinos than for higher-energy ones. This is because for a given field
strength, as the neutrino energy increases, more and more 
Landau levels become available for the electron in the final state
(See Eq.~[\ref{nmax}]). Consequently, the electron phase space and 
hence the cross section
should approach the zero-field values according to the correspondence
principle. We found that typically when the number of Landau levels filled
by the electron ($n_{\rm max}$) is greater than 2--3, 
the magnetic field effect
on the absorption opacity becomes negligible. 
The ``critical'' neutrino energy is approximately given by 
\be
E_\nu^{\rm (crit)}\simeq m_e\,(2bn_{\rm max}+1)^{1/2}
\simeq 6\left({n_{\rm max}\over 3}\right)^{1/2}
\left({B\over 10^{15}\,{\rm G}}\right)^{1/2}~
{\rm MeV}.
\ee
For $E_\nu\go E_\nu^{\rm (crit)}$, 
one can approximate $\langle\sigma^\abs_B(E_\nu)\rangle$
by the zero-field value $\langle\sigma^\abs_{B=0}(E_\nu)\rangle$
with good accuracy (within a few percent). 
Alternatively, we can define a threshold magnetic field, $B_{\rm th}$,
above which the $\nu_e$ and $\bar\nu_e$ absorption opacities 
are modified (by more than a few percent) by the magnetic field:
\be
B_{\rm th}\simeq {\langle E_\nu\rangle^2\over 2n_{\rm max}m_e^2}B_c
\simeq 3\times10^{15}\left({\langle E_\nu\rangle\over 10\ {\rm MeV}}\right)^2
\left({3\over n_{\rm max}}\right)\ {\rm G},
\ee
where $\langle E_\nu\rangle\sim 10$ MeV is the average neutrino energy.

In the deeper layers of the neutron star, the typical neutrino 
energy is higher. Therefore we expect the magnetic field effect 
on the absorption opacities to be smaller there. In addition, 
for $\nu_e$, absorption becomes unimportant as compared to scattering
in the high density region (although not necessarily in the deepest interior,
where $\nu_e$ can be degenerate). Thus, in order to assess the 
importance of the magnetic field effect on the neutrino emission, we only need
to focus on the surface region, the region between the neutrinosphere and
the neutrino-matter decoupling layer. We note that because the
absorption opacities in the magnetic field for different neutrino energies
oscillate around the the zero-field values,
without a detailed calculation of the mean opacity
(appropriately averaged over the neutrino energy spectrum) one cannot even 
answer the qualitative question such as whether the magnetic field increases or
decreases the local neutrino energy flux.  

In \S\S3--6, we study how the modified absorption cross
sections affect the emergent $\nu_e$ and $\bar\nu_e$ energy fluxes and spectra 
from the proto-neutron star.

\section{Neutrino Transport Near the Stellar Surface}

As discussed in \S2.4, the quantization effect of magnetic fields
on neutrino absorption is important in the region only near the neutron star 
surface (between the neutrinosphere and the neutrino-matter decoupling layer). 
In this section we consider neutrino transport in this
relatively low density region. 

Because $\nu_e$ and $\bar\nu_e$ with typical
energies of 10 MeV have comparable absorption and scattering opacities 
(see Fig.~3), the emergent $\nu_e$ and $\bar\nu_e$ energy fluxes 
and spectra are determined essentially in the same region
where these two neutrino species just begin to thermally
decouple from the proto-neutron star matter. The $\nu_e$
and $\bar\nu_e$ occupation numbers in this region are then
approximately given by
\begin{equation}
f_{\nu_e}(E_\nu)=
{1\over\exp[(E_\nu/T)-\eta_\nu]+1},
\label{occnue}\end{equation}
and
\begin{equation}
f_{\bar\nu_e}(E_\nu)=
{1\over\exp[(E_\nu/T)+\eta_\nu]+1},
\label{occnueb}\end{equation}
respectively, where $\eta_\nu=\mu_\nu/T$, and $\mu_\nu$ is
the $\nu_e$ chemical potential. From these occupation numbers, we
obtain
\begin{equation}
n_{\nu_e}-n_{\bar\nu_e}={\eta_\nu\over6}\left(1+{\eta_\nu^2\over\pi^2}\right)T^3,
\label{nuno}\end{equation}
and
\begin{equation}
U_{\nu_e}+U_{\bar\nu_e}={7\pi^2\over120}
\left(1+{30\eta_\nu^2\over7\pi^2}+{15\eta_\nu^4\over7\pi^4}\right)T^4,
\label{nuen}\end{equation}
where for example, $n_{\nu_e}$ and $U_{\nu_e}$ are the number and
energy densities, respectively, for the $\nu_e$.
We can estimate the magnitude of $\eta_\nu$ as follows.
The net deleptonization rate of the proto-neutron star is
\begin{equation}
-\dot N_e\sim{Y_{e,i}M/m_N\over t_{\rm diff}}
\sim c(n_{\nu_e}-n_{\bar\nu_e})\pi R^2,
\label{ratlep}\end{equation}
where $Y_{e,i}$ is the initial electron fraction of the proto-neutron star,
$M$ is the proto-neutron star mass, $m_N$ is the nucleon mass, 
$t_{\rm diff}$ is the neutrino diffusion timescale, and $R$ is the
proto-neutron star radius. The sum of the $\nu_e$ and $\bar\nu_e$
luminosities is
\begin{equation}
L_{\nu_e}+L_{\bar\nu_e}\sim{1\over 3}{GM^2/R\over t_{\rm diff}}
\sim c(U_{\nu_e}+U_{\bar\nu_e})\pi R^2.
\label{nulum}\end{equation}
From Eqs.~(\ref{nuno})--(\ref{nulum}), we have
\begin{equation}
\eta_\nu\sim{21\pi^2\over 20}{1+(30/7\pi^2)\eta_\nu^2+(15/7\pi^4)\eta_\nu^4
\over 1+(\eta_\nu^2/\pi^2)}{Y_{e,i}TR\over GMm_N}
\sim0.1\left({Y_{e,i}\over 0.36}\right)\left({T\over 5\ {\rm MeV}}\right)
\ll 1,
\label{etanu}\end{equation}
where the numerical value is obtained for 
$M\simeq 1.4M_\odot$ and $R\simeq 10$ km. Therefore, to a good approximation,
the $\nu_e$ and $\bar\nu_e$ in the decoupling region have Fermi-Dirac energy
distributions with zero chemical potentials. 

Let $I_E=I_E(\br,\hato)$ be the specific intensity of
neutrinos of a given species, where $\br$ is the spatial 
position, and $\hato$ specifies the direction of propagation. 
The general transport equation takes the form
\be
\hato\cdot\nabla I_E=\rho\kappa_E^{\abs\ast}I_E^{(FD)}
+\rho\kappa^\sc_E {1\over 4\pi}\int\!d\Omega' I_E'
-\rho\, \left[\kappa_E^{\abs\ast}+\kappa_E^\sc\right]\,I_E,
\label{Inu}\ee
where $I_E'=I_E(\br,\hato')$, and
the subscript ``$E$'' indicates that 
the relevant physical quantities depend on the neutrino energy $E$.
The opacities $\kappa_E^{\abs\ast}$ and $\kappa_E^\sc$
are related to the corresponding cross sections as, e.g.,
$\kappa_E^{\abs\ast}=(1-Y_e)\sigma_E^{\abs\ast}/m_N$ and
$\kappa_E^\sc=\sigma_E^\sc/m_N$ for $\nu_e$. Note that we have 
included the effect of stimulated absorption for neutrinos 
(e.g., Imshennik \& Nadezhin 1973) by introducing the corrected
absorption opacity $\kappa_E^{\abs\ast}=\kappa_E^\abs
[1+\exp(-E/T)]$ ($T$ is the matter temperature). 
Hereafter, we shall suppress the superscript ``$\ast$''
in $\kappa_E^{\abs\ast}$, and the notation $\kappa_E^\abs$
should be understood as having included the correction due to
stimulated absorption.
The quantity $I_E^{(FD)}$ in Eq.~(\ref{Inu}) is the equilibrium 
neutrino intensity, given by the Fermi-Dirac distribution
\be
I_E^{(FD)}={E^3\over c^2h^3}{1\over \exp(E/T)+1}.
\ee
Note that the time derivative term $(\partial I_E/\partial t)$
on the left-hand-side of Eq.~(\ref{Inu}) has been neglected.
This corresponds to an instantaneous redistribution of emission and
absorption sources (i.e., an instantaneous 
redistribution of matter temperature)
--- The timescale for the redistribution is of order the mean free path
divided by $c$, much smaller than the neutrino cooling time of the star.

The zeroth order moment of the transport equation can be written as
\be
\nabla\cdot\bF_E=\rho\,\kappa_E^{\abs} c\,(U_E^{(FD)}-U_E),
\label{dfnu}\ee
where $\bF_E=\int\!I_E\hato d\Omega$ is the (spectral) neutrino
energy flux, $U_E=\int\! I_E d\Omega/c$ is the energy density, and
$U_E^{(FD)}=(4\pi/c)I_E^{(FD)}$ is the corresponding equilibrium energy
density. With the Eddington closure relation, 
the first moment of Eq.~(\ref{Inu}) becomes  
\be
\bF_E=-{c\over 3\,\rho\,\kappa_E^{(t)}}\nabla U_E,
\label{fnu}\ee
where $\kappa_E^{(t)}=\kappa_E^{\abs}+\kappa_E^\sc$.
Equations (\ref{dfnu}) and (\ref{fnu}) completely specify
the neutrino radiation field. 

\section{Rosseland Mean Opacities}

Here we are concerned with the total neutrino energy flux emitted 
from a local surface region of the neutron star. The magnetic field
strength in this local region is assumed to be constant. 
Since the scale heights for physical
quantities such as the matter density and temperature are much smaller
than the local radii, we adopt the plane-parallel geometry. 
Integrating Eq.~(\ref{fnu}) over the neutrino energy, we find that the
total neutrino energy flux (of a specific species) 
$\bF=\int\bF_E dE=F {\hat{\bf r}}$ is given by 
\be
F=-{c\over 3\,\rho\,\kappa_R}{dU\over dr},
\label{ftot}\ee
where $U=\int U_EdE$ is the total neutrino energy density, and
$\kappa_R$ is the ``Rosseland mean opacity'' defined as
\be
{1\over\kappa_R}={\int\!dE\,({1/\kappa_E^{(t)}})\,({d
U_E/dr})\over\int\!d E\,({dU_E/dr})}.
\label{ross}\ee
Since neutrino emission and absorption rates are extremely
fast even in the regions near the proto-neutron star surface, a radiative
equilibrium\footnote{In principle, the overall radiative equilibrium
involves matter and neutrinos of all species. However, because the coupling
between matter and $\nu_\mu$, $\bar\nu_\mu$, $\nu_\tau$, and $\bar\nu_\tau$
is much weaker, the overall radiative equilibrium in the region near the
proto-neutron star surface mainly involves matter and $\nu_e$ and $\bar\nu_e$.
Since the coupling between matter and $\nu_e$ is comparable to that between
matter and $\bar\nu_e$, we further assume that 
an approximate radiative equilibrium
holds separately for $\nu_e$ and $\bar\nu_e$.}
approximately holds, i.e.,
\be
\int\rho\,\kappa_E^{\abs} (U_E^{(FD)}-U_E)dE\simeq 0.
\ee
>From Eq.~(\ref{dfnu}), we have $dF/dr\simeq 0$, or $F$ is approximately
constant. 

If we introduce the Rosseland mean optical depth measured from the stellar
surface via $d\tau_R=-\rho\kappa_R dr$, with $F$ constant,
Eq.~(\ref{ftot}) can be integrated to yield
\be
cU=3F\left(\tau_R+{2\over 3}\right),
\label{profile}\ee
where we have used the boundary condition $F=cU/2$ at the surface
($\tau_R=0$). Equation (\ref{profile}) resembles the standard Eddington
profile. 
Note that outside the decoupling region the neutrino energy density 
U may not be equal to $U^{(FD)}=(7\pi^4/240)T^4$ specified by the matter
temperature $T$. Similarly, $U_E$ may differ substantially 
from $U_E^{(FD)}(T)$ (see \S6). However, 
we can still approximate $U_E$ by the Fermi-Dirac function at 
the neutrino temperature $T_\nu$, which is equal to the matter temperature
at the decoupling layer. 
Equation (\ref{ross}) can then be rewritten as
\be
{1\over\kappa_R}=
{30\over 7\pi^4T_\nu^5}\int\!dE{1\over\kappa_E^{(t)}}
{E^4 \exp (E/T_\nu)\over [\exp(E/T_\nu)+1]^2}.
\label{rosst}\ee
With the cross sections given in \S2, we can calculate the Rosseland
mean opacity $\kappa_R(B)$ in magnetic fields using Eq.~(\ref{rosst}).
Note that in Eq.~(\ref{rosst}) $\kappa_E^{(t)}$ depends on the local matter 
temperature $T$. Only in the regions not far from the decoupling layer, 
do we have $T_\nu=T$. 

Figure 4 shows the ratio of the Rosseland mean opacity $\kappa_R(B)$ of 
$\nu_e$ to the zero-field value $\kappa_R(0)$ as a function of the magnetic
field strength for typical conditions near the proto-neutron star surface.
We choose these conditions to be 
$T=T_\nu=3,~5$~MeV, and $\mu_e=10,~15,~20$~MeV, 
all with $Y_e=0.1$. Figure 5 shows similar results for $\bar\nu_e$ 
with similar physical parameters ($T=T_\nu=3,~5,~7$~MeV; Note that the 
results are insensitive to $\mu_e$). We see that 
deviation of $\kappa_R(B)$ from the zero-field value by more than 
$5\%$ generally requires field strength $b=B/B_c\go 100$. 

The behavior of the ratio $\kappa_R(B)/\kappa_R(0)$ as a function of
$b$ is certainly not obvious a priori. Since the Rosseland mean integral 
is dominated by neutrinos with the smallest opacities, and since 
$\kappa^\abs_B$ scales linearly with $B$ in the regime where 
only the ground Landau level is filled (see Eq.~[\ref{sigbapp}]), 
we expect that for sufficiently large $B$, the Rosseland mean opacity 
increases linearly with increasing $B$. This is indeed what is indicated
in Figs.~4 and 5. For ``intermediate'' field strength, the $\kappa^\abs_B$
curve for $\nu_e$ (see Fig.~1) has a significant portion which is below 
the zero-field value at low neutrino energies. For such intermediate 
fields, we would expect $\kappa_R(B)$ to be less than the zero-field
value. This explains the non-monotonic behavior of $\kappa_R(B)/\kappa_R(0)$
as a function of $B$ shown in Fig.~4. For $\bar\nu_e$, 
the non-monotonic behavior is much less pronounced (see the 
curve for $T=7$ MeV in Fig.~5) since the energy range in which $\kappa_B^\abs
<\kappa_{B=0}^\abs$ is smaller.

\section{Differential Neutrino Fluxes and Neutron Star Kick 
>From Asymmetric Magnetic 
Field Topology}

Imagine a proto-neutron star with an asymmetric magnetic field
topology, e.g., the north pole and the south pole have different
{\it magnitudes} of magnetic field (say, by a factor of a few). 
What is the difference in the neutrino fluxes from these two regions
due to the different absorption opacities of $\nu_e$ and $\bar\nu_e$,
and what is the resulting kick velocity of the neutron star? We
recall that the fractional asymmetry $\alpha$ in the total neutrino emission
(of all species) required to generate a kick velocity $V_{\rm kick}$
is $\alpha=0.028\,[V_{\rm kick}/(10^3~{\rm km~s}^{-1})]
\,[E_{\rm tot}/(3\times 10^{53}~{\rm erg})]$,
where $E_{\rm tot}$ is the total neutrino energy radiated from the neutron
star (the neutron star mass has been set to $1.4\,M_\odot$). 

To quantitatively understand how the 
magnetically-modified opacities change the 
local neutrino flux, we will have to evaluate these opacities
at different depths (corresponding to different physical conditions) below the
neutron star surface, and to perform complete cooling calculations which take
into account these opacities. However, 
Figs.~4 and 5 clearly indicate that, even near the stellar surface,  
where one expects the modification to the opacities to be most prominent, 
and even with the most favorable conditions for the magnetic field
to be effective in changing the opacities (i.e., low density and low
temperature), a $5\%$ change in the Rosseland mean opacity (for $\nu_e$ or
$\bar\nu_e$) requires a magnetic field with $b\go 100$, or
$B\go 4\times 10^{15}$~G. 
In the following, we shall content ourselves with merely estimating
the change in the local neutrino flux from the modified opacities.

We first consider the case in which the matter temperature throughout
the neutron star is unaffected by the magnetic field. One might imagine this
as a possibility if convection dominates energy transfer between 
different parts of the star. Under the neutrino-matter decoupling layer
the specific neutrino energy density $U_E$ is the same as the Fermi-Dirac 
function $U_E^{(FD)}(T)$ specified by the matter temperature. 
The emergent neutrino flux is then inversely proportional to 
the Rosseland mean optical depth $\tau_R$ from the decoupling layer to the
stellar surface (see Eq.~[\ref{profile}]). Since $\tau_R$ for $\nu_e$ or
$\bar\nu_e$ at the decoupling layer is of order a few, the flux $F$ 
is inversely proportional to $\kappa_R$ evaluated near the decoupling region. 
If we use the curve for $T=3$~MeV and $\mu_e=10$~MeV in Fig.~4 
(which has the most 
pronounced magnetic correction), we find that the fractional 
change in the $\nu_e$ flux is $\delta F_\nue/F_\nue\sim 0.1$ at $b=100$
(If we used the curve for $\mu_e=15$ or 20~MeV, which more closely
represents the physical condition at the decoupling region, the modification
$\delta F_\nue/F_\nue$ would be smaller by a factor of a few or more). 
Similarly, using the curve for $T=3$~MeV in Fig.~5 
to maximize the magnetic correction,
we find $\delta F_\bnue/F_\bnue\sim -0.02$ at $b=100$. Since the fluxes
of other types of neutrinos are unchanged\footnote{The decoupling
layers of $\nu_{\mu(\tau)}$ and $\bar\nu_{\mu(\tau)}$ depend on the $\nu$-$e$
scattering cross sections, which can be affected by the magnetic field.
However, the layers are located much deeper than those of
$\nu_e$ and $\bnue$ (since $\nu$-$e$ scattering opacity is small),
the modification to the $\nu$-$e$ scattering opacity due to 
electron Landau levels is therefore smaller than the modification to the
absorption opacities of $\nue$ and $\bnue$ considered in this paper.},
and since different neutrino species more or less carry away the same 
amount of energy, the change in the total neutrino flux is only
$\delta F_{\rm tot}/F_{\rm tot}\sim 0.01$ at $b=100$. If we now imagine that
the north pole of the neutron star has $b\sim 100$ while the south pole
has a field strength a factor of a few smaller, then, taking into account 
the geometric reduction ($\sim 1/3$), the flux asymmetry is about
$\alpha\sim 0.003$, just enough to give a kick velocity of 
$\sim 100$~km~s$^{-1}$ to the neutron star.
With $\kappa_R$ evaluated at more realistic conditions, we expect 
the resulting $\alpha$ to be even smaller. 

The assumption that the matter temperature is unaffected by the magnetic 
field (and therefore spherically symmetric) is unlikely to be valid. 
In reality, when a region of the star (say the north pole) emits
more flux, it also cools faster. With reduced temperature, this region 
will tend to emit less flux. In the extreme case, one would find
$F\sim cU_0/\tau_R$, with $U_0$ 
the neutrino energy density in the stellar core,
and $\tau_R$ the Rosseland mean optical depth from the core to the surface.
Since in the bulk interior of the star, modification to $\kappa_R$ 
due to the magnetic field is negligible, we would conclude that 
the change in the neutrino flux is much smaller than what is estimated
in the last paragraph. 

To summarize, the modifications to the absorption opacities of 
$\nu_e$ and $\bnue$ due to quantized Landau levels give rise to
rather small change in neutrino fluxes from the neutron star. 
To generate appreciable kick velocities ($\sim 300$~km~s$^{-1}$)
from this effect requires the difference of field strengths
at the two opposite poles to be at least $10^{16}$~G (and possibly
much larger).

\section{Magnetic Field Effect on $\nu_e$ and $\bar\nu_e$ Energy Spectra}

We now consider how the oscillatory behaviors (see Figs.~1--3)
of the absorption opacities of $\nu_e$ and $\bar\nu_e$ may manifest 
themselves in the radiated neutrino energy spectra. 
For an absorption dominated medium, neutrinos with energy
$E$ decouple from the matter and start free-streaming
at a column depth (in g~cm$^{-2}$) $y=\int\!\rho\,dr\sim 1/\kappa_E^{\abs}$.
Thus the neutrino energy spectrum is given by $F_E\sim cU_E^{(FD)}$ with
$U_E^{(FD)}$ evaluated at the depth $y\sim 1/\kappa_E^{\abs}$.
In the presence of significant scattering, decoupling occurs
at depth $y\sim 1/\kappa_E^{\rm eff}$ with the effective
opacity $\kappa_E^{\rm eff}
\sim\sqrt{\kappa_E^{\abs}\kappa_E^{(t)}}$, whereas
the neutrinosphere from which neutrinos start free-streaming
is located at $y\sim 1/\kappa_E^{(t)}$.
Near the decoupling region we have
$dU_E/dr\sim -\rho\kappa_E^{\rm eff}U_E^{(FD)}$, and thus
from Eq.~(\ref{fnu}) the neutrino spectral energy flux is given by 
\be
F_E\sim {\kappa_E^{\rm eff}\over\kappa_E^{(t)}}
\,c\,U_E^{(FD)}(y\sim 1/\kappa_E^{\rm eff}).
\ee

The above qualitative argument can be made more rigorous by 
directly solving Eqs.~(\ref{dfnu}) and (\ref{fnu}). Consider
a plane-parallel atmosphere and define the (energy-dependent) optical
depth via $d\tau_E=-\rho\kappa_E^{(t)}dr=
\kappa_E^{(t)} dy$. Equations (\ref{dfnu}) and (\ref{fnu}) can be
combined to give 
\be
{d^2\over d\tau_E^2}U_E=\beta_E^2\, [U_E-U_E^{(FD)}],~~~~~
{\rm with}~~~\beta_E^2=3\,{\kappa_E^{\abs}\over\kappa_E^{(t)}}.
\ee
With a constant $\beta_E$,
the above equation can be solved exactly. 
For the boundary condition $F_E=(c/3)(dU_E/d\tau_E)=cU_E/2$ at $\tau_E=0$,
the neutrino energy flux at the surface 
is given by
\be
F_E={\beta_E/3\over 1+2\beta_E/3}\int_0^\infty
\!d\tau_E\,\beta_E\,c\,U_E^{(FD)}(\tau_E)\,\exp(-\beta_E\tau_E)
\simeq {\beta_E/3\over 1+2\beta_E/3}\,
c\,U_E^{(FD)}(\tau_E\simeq\beta_E^{-1}).
\label{fnu2}\ee

Figure 6 shows the spectral energy fluxes of $\nu_e$
calculated using Eq.~(\ref{fnu2}) for different field strengths ($b=100$ and
$300$).  The matter temperature profile is
obtained from Eq.~(\ref{profile}) assuming $U\propto T^4$.
For a total energy flux of $F=(c/4)U(T_{\rm eff})$ (where $T_{\rm eff}$ 
is the effective temperature), we have $T=T_{\rm eff}[(3/4)\tau_R+1/2]^{1/4}$.
We choose $T_{\rm eff}=3$ MeV in these examples. For simplicity, we have
neglected the variations of the opacities as a function of $\tau_E$. Thus the 
temperature at decoupling ($\tau_E\simeq\beta_E^{-1}$; see Eq.~[\ref{fnu2}])
is given by $T\simeq T_{\rm eff}
[(3/4)\kappa_R/\kappa_E^{(\rm eff)}+1/2]^{1/4}$, where
$\kappa_E^{(\rm eff)}=\beta_E\kappa_E^{(t)}=\sqrt{3\kappa_E^\abs
\kappa_E^{(t)}}$ 
and we take the 
relevant opacities to have the corresponding values for $Y_e=0.1$,
$\mu_e=20$ MeV, and $T=3$ MeV. 
This simplification should not introduce qualitative changes
in the spectrum. As seen from Fig.~6, the spectrum is pinched
at the high energies because of the $E^2$-dependence of the opacities. 
The magnetic field introduces modulations in the spectrum. For lower
magnetic field strengths, the modulations occur at the lower energy band.
As the magnetic field gets stronger, the modulations shift to the higher 
energy part of the spectrum. 

The distortions and modulations of the $\nu_e$ and 
$\bar\nu_e$ energy spectra by strong magnetic fields may offer
an interesting possibility to limit the magnetic field strength
in proto-neutron stars through the detection of neutrinos from
a Galactic supernova. We note, however, that the $\nu_e$ spectra 
shown in Fig.~6 are based on very approximate calculations. They serve only as
illustrative examples of the magnetic field effect on the neutrino spectra.
More detailed calculations are needed to obtain realistic spectra
which include the magnetic field effect.

\section{Conclusion}

In this paper, we have studied the issue of whether asymmetric 
magnetic field topology in a proto-neutron star can induce
asymmetric neutrino emission from the star through the
modifications of the neutrino absorption opacities by the magnetic 
field. These modifications arise from the 
quantized Landau levels of electrons and positrons produced in strong magnetic 
fields. By calculating the appropriate mean neutrino opacities
in the magnetic field, we demonstrate that this mechanism is rather 
inefficient in generating a kick to the neutron star: To obtain
appreciable kick velocities ($\sim 300$~km~s$^{-1}$), the difference in 
the field strengths at the two opposite poles of the star must be at least
$10^{16}$~G. 

\acknowledgments

This work was started while D.L. held the Richard C. Tolman 
postdoctoral Fellowship at Caltech. Additional support was provided
by NASA grant NAG 5-2756 and NSF grant AST-9417371, as well as
a start-up fund for new faculty at Cornell University.
Y.Q. is supported by the David W. Morrisroe postdoctoral Fellowship
at Caltech.


\bigskip

\begin{figure}
\plotone{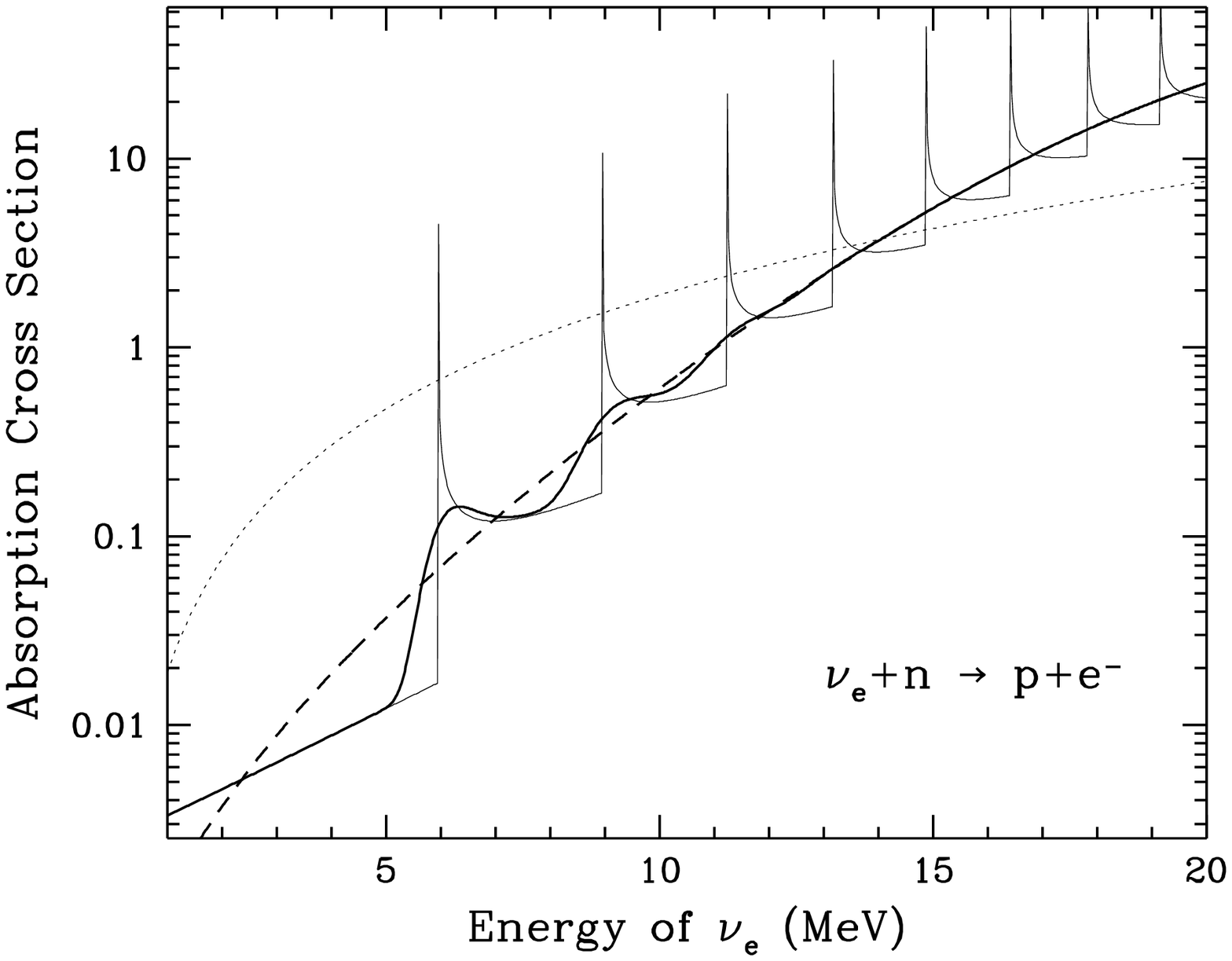}
\caption{
Absorption cross section (per nucleon) of $\nu_e$, 
$(1-Y_e)\sigma_B^\abs(E_\nu)$, 
in units of $A\times{\rm MeV}^2=\pi G_F^2\cos^2\theta_C(c_V^2+3c_A^2)
\times{\rm MeV}^2\simeq 
0.91\times10^{-42}$~cm$^2$, as a function
of the neutrino energy. The dashed line is the zero-field result
(The effect of the nucleon thermal motion is included but is hard to discern).
The heavy and the light solid lines are the results for $b=100$,
with and without the effect of the nucleon thermal motion, respectively. 
The dotted line is the elastic scattering cross section, $\sigma^\sc(E_\nu)$,
in the same units. 
All results are for $T=3$~MeV, $Y_e=0.1$, and $\mu_e=20$~MeV 
(corresponding to $\rho=7.2\times 10^{11}$ g$\,$cm$^{-3}$).
}
\end{figure}

\begin{figure}
\plotone{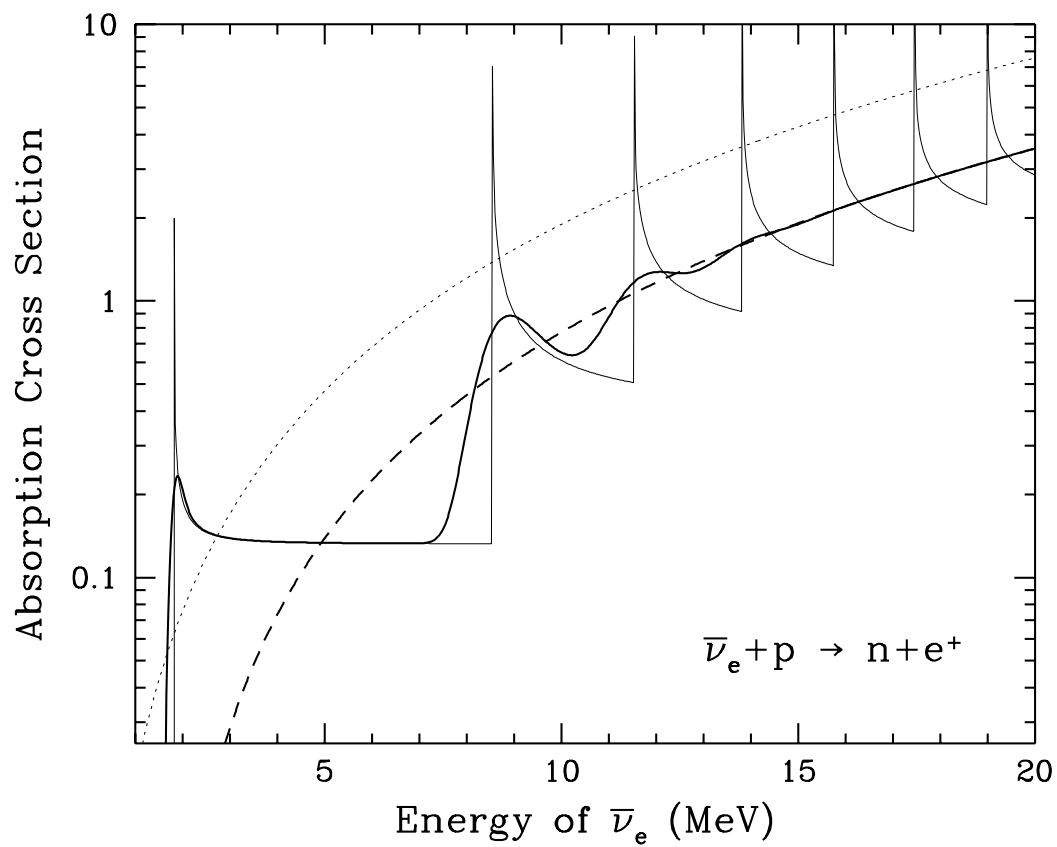}
\caption{
Absorption cross section (per nucleon) of $\bar\nu_e$,
$Y_e\sigma_B^\abs(E_\nu)$. 
The physical parameters, units, and labels are the same as in Fig.~1. 
}
\end{figure}

\begin{figure}
\plotone{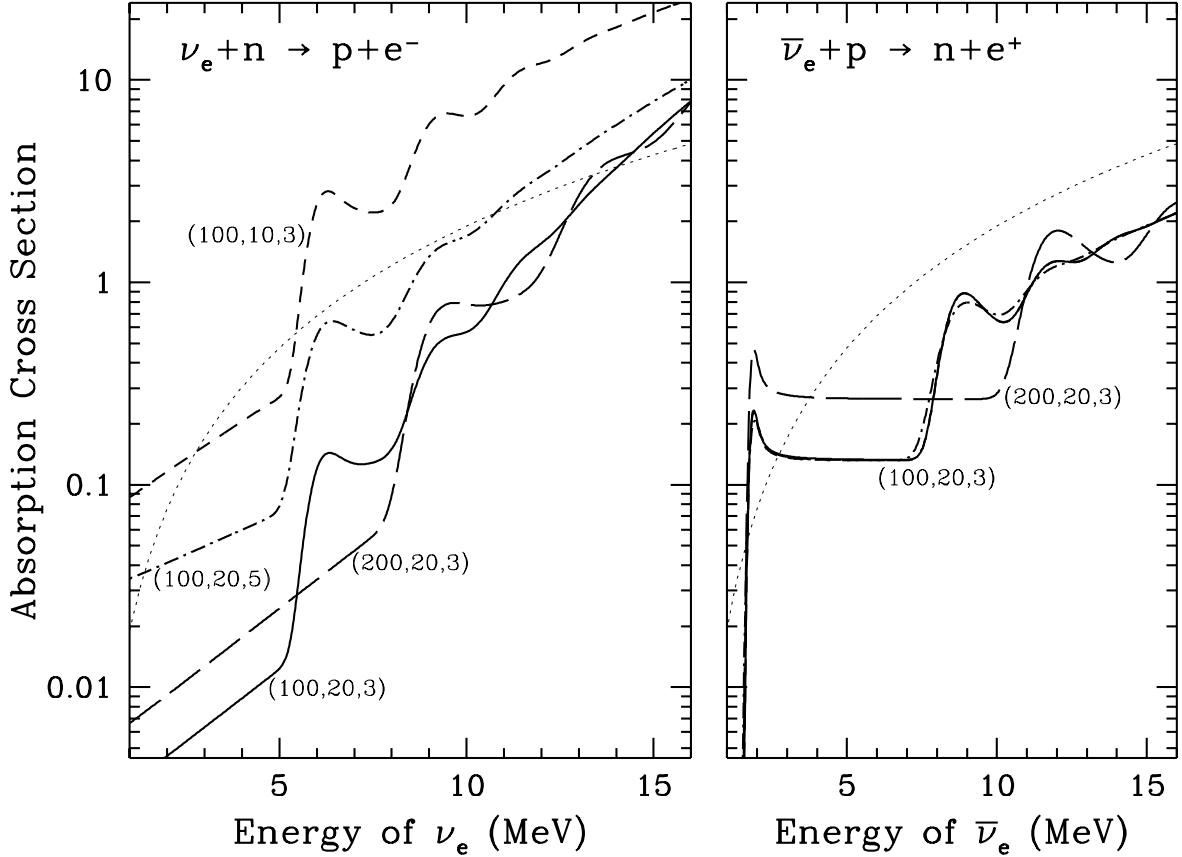}
\caption{
Absorption cross sections (per nucleon) of $\nu_e$ (left panel) and
$\bar\nu_e$ (right panel), 
in units of $A\times{\rm MeV}^2=\pi G_F^2\cos^2\theta_C(c_V^2+3c_A^2)
\times{\rm MeV}^2\simeq
0.91\times10^{-42}$~cm$^2$, as functions of the neutrino energy for different 
physical parameters. All results assume $Y_e=0.1$. The solid, 
the short-dashed, the long-dashed, and the dot-dashed curves
are for $(b,\mu_e/{\rm MeV},T/{\rm MeV})=(100,20,3)$, (100,10,3),
(200,20,3), and (100,20,5), respectively.
[Note that for $\bar\nu_e$, the curve for 
$(b,\mu_e/{\rm MeV},T/{\rm MeV})=(100,10,3)$
is almost identical 
to the curve for $(b,\mu_e/{\rm MeV},T/{\rm MeV})=(100,20,3)$, 
and both curves are close
to the curve for $(b,\mu_e/{\rm MeV},T/{\rm MeV})=(100,20,5)$.]
The light dotted lines are the elastic scattering cross sections
in the same units. 
}
\end{figure}

\begin{figure}
\plotone{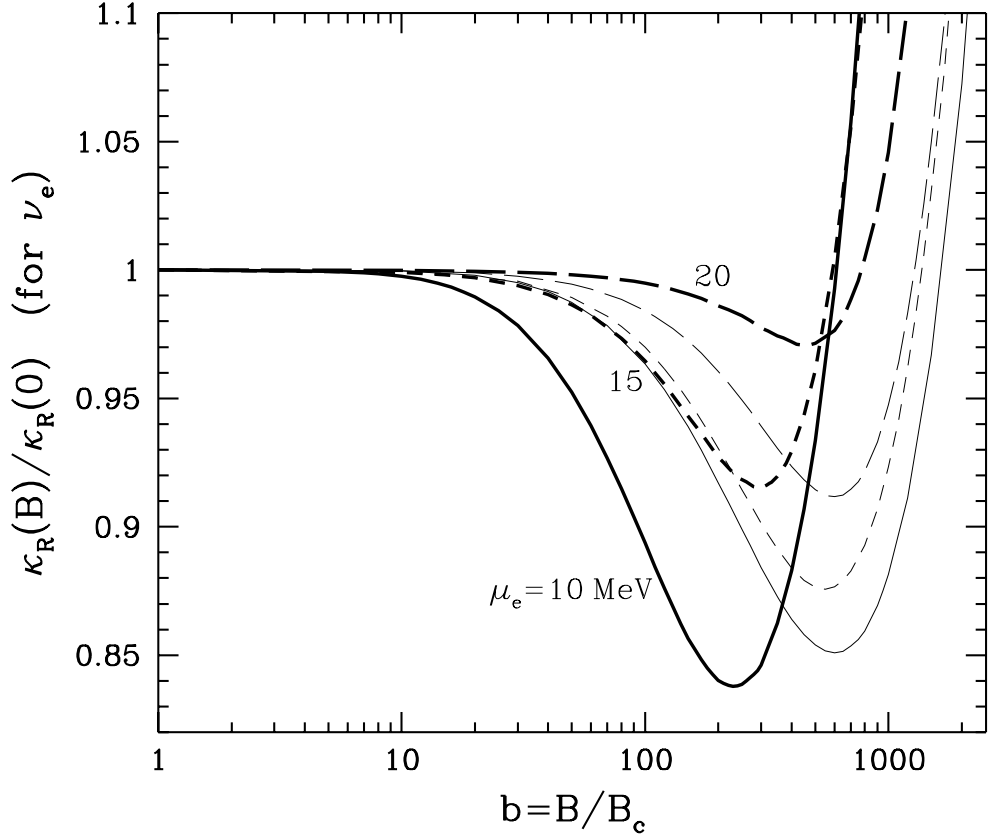}
\caption{
The ratio of the Rosseland mean opacity $\kappa_R(B)$ of $\nu_e$ 
to its zero-field value $\kappa_R(0)$ as a function of the magnetic 
field strength $b=B/B_c$ (with $B_c=4.414\times 10^{13}$~G). 
All results assume $Y_e=0.1$. The heavier curves correspond 
to $T=T_\nu=3$~MeV, and the lighter cuvres correspond to $T=T_\nu=5$~MeV. 
The solid, the short-dashed, and the long-dashed lines are for 
$\mu_e=10$, 
15,
and 20~MeV, respectively.
}
\end{figure}

\begin{figure}
\plotone{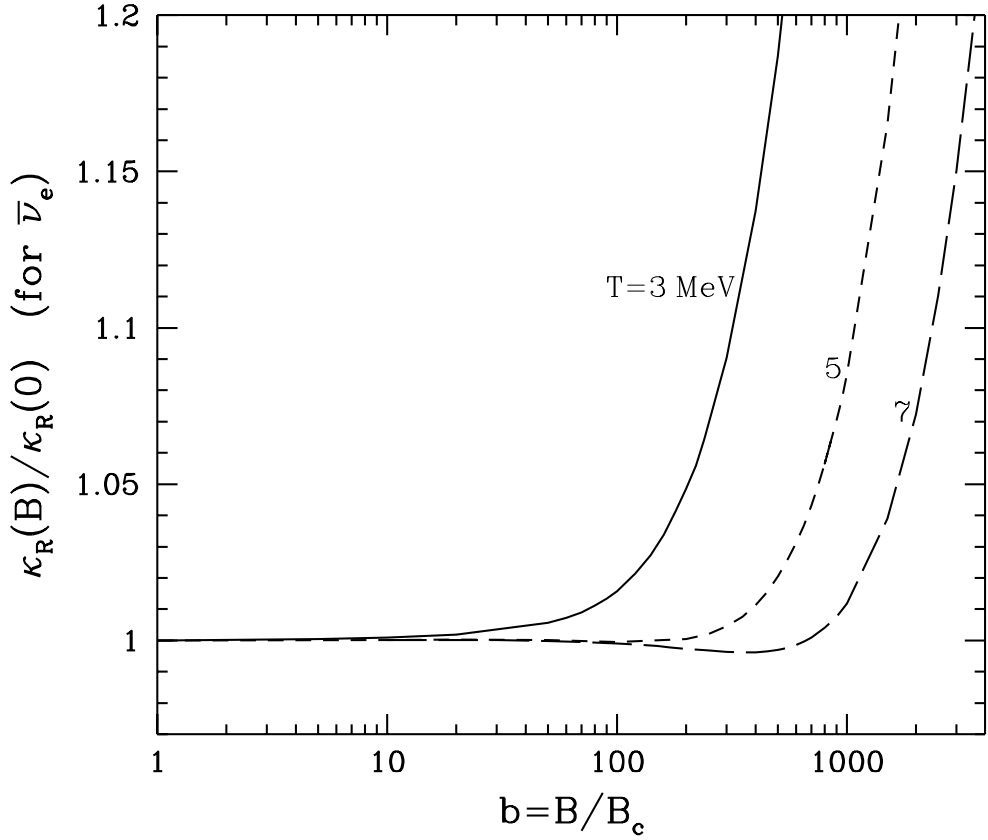}
\caption{
The ratio of the Rosseland mean opacity $\kappa_R(B)$ of $\bar\nu_e$ 
to its zero-field value $\kappa_R(0)$ as a function of the magnetic 
field strength $b=B/B_c$ (with $B_c=4.414\times 10^{13}$~G). 
All results assume $Y_e=0.1$ and $\mu_e=20$~MeV (but the results are rather
insensitive to the value of $\mu_e$).
The solid, the short-dashed, and the long-dashed line are for 
$T=T_\nu=3$, 
5, and 
7~MeV, respectively. Note that the results mainly depend on $T_\nu$, and are 
rather insensitive to $T$.
}
\end{figure}

\begin{figure}
\plotone{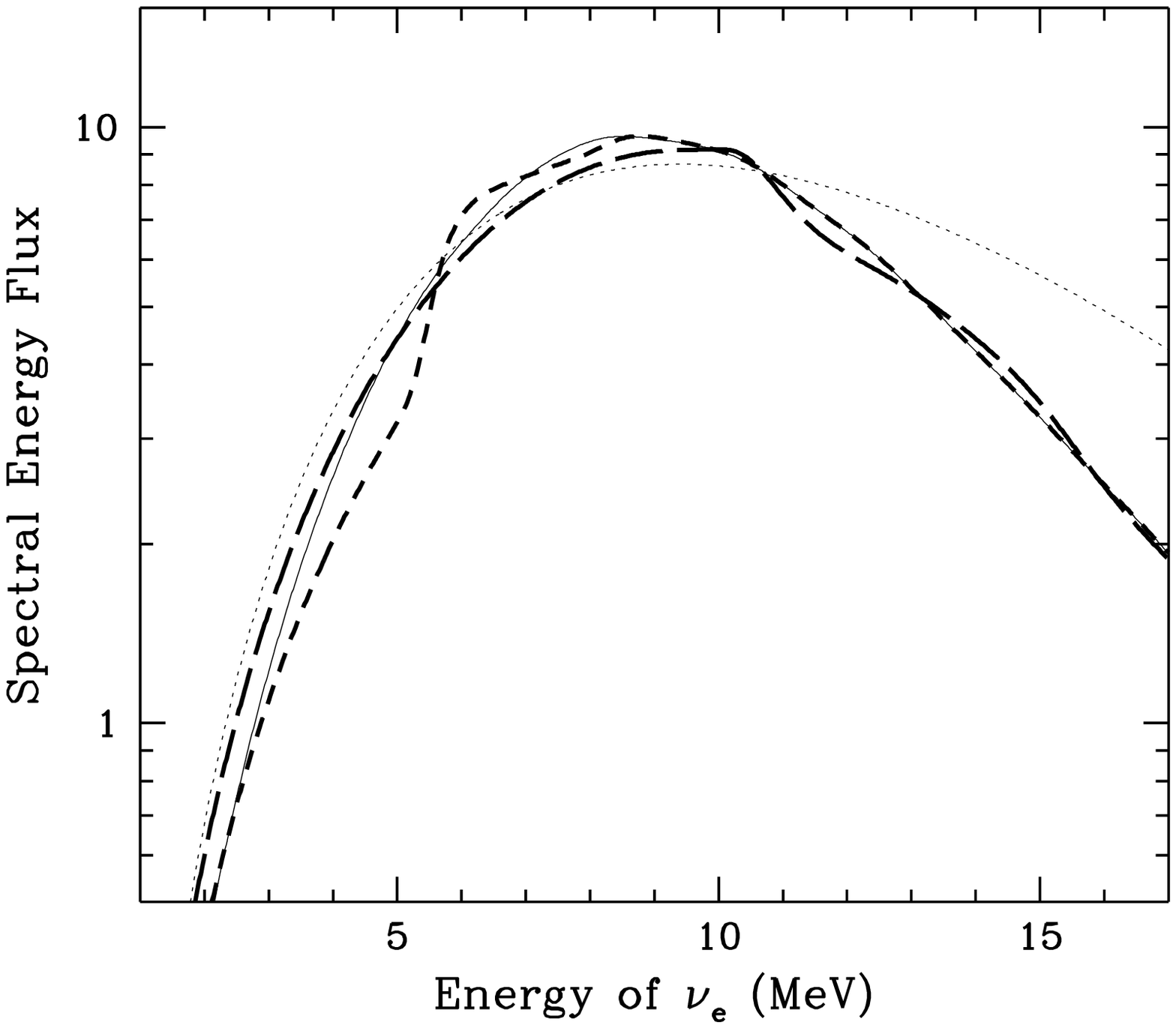}
\caption{
Illustrative examples of the $\nu_e$ spectral energy fluxes 
(in arbitrary units)
modified by the magnetic fields. The light solid, the short-dashed,
and the long-dashed line are for $b=0$, 100, and 300, respectively.
The light dotted line is the Fermi-Dirac spectrum
with $T=T_{\rm eff}=3$ MeV.
}
\end{figure}


\begin{thebibliography}{}

\bibitem{}
Akhmedov, E.~K., Lanza, A., \& Sciama, D.~W. 1997, Phys. Rev. D, 56, 6117.

\bibitem{}
Arras, P., \& Lai, D. 1998, submitted to Phys. Rev. Lett.

\bibitem{}
Arzoumanian, Z., Phillips, J.~A., Taylor, J.~H., \& Wolszczan, A. 1996,
ApJ, 470, 1111.

\bibitem{}
Aschenbach, B., Egger, R., \& Trumper, J. 1995, Nature, 373, 587.

\bibitem{}
Bazan, G., \& Arnett, D. 1994, ApJ, 433, L41.

\bibitem{}
Bezchastnov, V.~G., \& Haensel, P. 1996, Phys. Rev. D, 54, 3706.

\bibitem{}
Bisnovatyi-Kogan, G.~S. 1993, Astron. Astrophys. Trans., 3, 287.

\bibitem{}
Brandt, N., \& Podsiadlowski, P. 1995, MNRAS, 274, 461.

\bibitem{}
Burrows, A., \& Fryxell, B.~A. 1992, Science, 258, 430.

\bibitem{}
Burrows, A., Hayes, J., \& Fryxell, B.~A. 1995, ApJ, 450, 830.

\bibitem{}
Burrows, A., \& Hayes, J. 1996, Phys. Rev. Lett., 76, 352.

\bibitem{}
Chugai, N.~N. 1984, Sov. Astron. Lett., 10, 87.

\bibitem{Cordes97}
Cordes, J.~M., \& Chernoff, D.~F. 1997, ApJ, submitted (astro-ph/9707308).

\bibitem{}
Cordes, J.~M, Romani, R.~W., \& Lundgren, S.~C. 1993, Nature, 362, 133.

\bibitem{}
Cordes, J.~M., Wasserman, I., \& Blaskiewicz, M. 1990, ApJ, 349, 546.

\bibitem{}
Cropper, et al. 1988, MNRAS, 231, 695.

\bibitem{}
Deway, R.~J., \& Cordes, J.~M. 1987, ApJ, 321, 780.

\bibitem{}
Dorofeev, O.~F., et al.~1985, Sov. Astron. Lett., 11, 123.

\bibitem{}
Fassio-Canuto, L. 1969, Phys. Rev., 187, 2141.

\bibitem{}
Frail, D.~A., Goss, W.~M., \& Whiteoak, J.~B.~Z.
1994, ApJ, 437, 781.

\bibitem{}
Fryer, C., Burrows, A., \& Benz, W. 1998, ApJ, in press (astro-ph/9710333)

\bibitem{}
Fryer, C., \& Kalogera, V. 1997, ApJ, in press (astro-ph/9706031).

\bibitem{}
Goldreich, P., Lai, D., \& Sahrling, M. 1996,
in ``Unsolved Problems in Astrophysics'', ed. J.~N. Bahcall and
J.~P. Ostriker (Princeton University press).

\bibitem{}
Gott, J.~R., Gunn, J.~E., \& Ostriker, J.~P. 1970, ApJ, 160, L91.

\bibitem{}
Gunn, J.~E., \& Ostriker, J.~P. 1970, ApJ, 160, 979.

\bibitem{}
Hansen, B.~M.~S., \& Phinney, E.~S. 1997, MNRAS, 291, 569.

\bibitem{}
Harrison, E.~R., \& Tademaru, E. 1975, ApJ, 201, 447.

\bibitem{}
Harrison, P.~A., Lyne, A.~G., \& Anderson, B.
1993, MNRAS, 261, 113.

\bibitem{}
Herant, M., Benz, W., Hix, W. R., Fryer, C. L., \& Colgate, S. A. 1994,
\apj, 435, 339.

\bibitem{}
Horowitz, C.~J., \& Li, G. 1997, astro-ph/9705126.

\bibitem{}
Horowitz, C.~J., \& Piekarewicz, J. 1997, hep-ph/9701214.

\bibitem{}
Iben, I., \& Tutukov, A.~V. 1996, ApJ, 456, 738.

\bibitem{}
Imshennik, V.~S., \& Nadezhin, D.~K. 1973, Sov. Phys. JETP, 36, 821.

\bibitem{}
Janka, H.-T., \& M\"uller, E. 1994, A\&A, 290, 496.

\bibitem{}
Janka, H.-T., \& M\"uller, E. 1996, A\&A, 306, 167.

\bibitem{}
Kaspi, V.~M., et al. 1996, Nature, 381, 584.

\bibitem{}
Kusenko, A., \& Segr\'e, G. 1996, Phys. Rev. Lett., 77, 4872.

\bibitem{}
Lai, D. 1996, ApJ, 466, L35.

\bibitem{}
Lai, D., Bildsten, L., \& Kaspi, V.~M. 1995, ApJ, 452, 819.

\bibitem{}
Lai, D., \& Goldreich, P. 1998, in preparation.

\bibitem{}
Lai, D., \& Qian, Y.-Z. 1998, \apjl, 495, L103.

\bibitem{Lorimer97}
Lorimer, D.~R., Bailes, M., \& Harrison, P.~A. 1997, MNRAS, 289, 592.

\bibitem{}
Lyne, A.~G., \& Lorimer, D.~R. 1994, Nature, 369, 127.

\bibitem{}
Matese, J.~J., \& O'Connell, R.~F. 1969, Phys. Rev., 180, 1289.

\bibitem{}
McCray, R. 1993, ARA\&A, 31, 175.

\bibitem{}
Minkowski, R. 1970, PASP, 82, 470.

\bibitem{}
Morse, J.~A., Winkler, P.~F., \& Kirshner, R.~P. 1995, AJ, 109, 2104.

\bibitem{}
Qian, Y.-Z. 1997, \prl, 79, 2750.

\bibitem{}
Roulet, E. 1997, Preprint, hep-ph/9711206.

\bibitem{}
Taylor, J.~H., \& Cordes, J.~M. 1993, ApJ, 411, 674.

\bibitem{}
Trammell, S.~R., Hines, D.~C., \& Wheeler,
J.~C. 1993, ApJ, 414, L21.

\bibitem{}
Tubbs, D.~L., \& Schramm, D.~N. 1975, ApJ, 201, 467.

\bibitem{}
Utrobin, V.~P., Chugai, N.~N., \& Andronova, A.~A. 1995, A\&A, 295, 129.

\bibitem{}
Vilenkin, A. 1995, ApJ, 451, 700.

\bibitem{}
Ziman, J.~M. 1979, Principles of the Theory of Solids (Cambridge Univ.
Press: Cambridge), p.~319.

\end{thebibliography}
\end{document}